# A high rate flow-focusing foam generator

Elise Lorenceau, Yann Yip Cheung Sang, Reinhard Höhler, Sylvie Cohen-Addad

*Université de Marne-la-Vallée*

*Laboratoire de Physique des Matériaux Divisés et des Interfaces, UMR 8108 du CNRS*

*5 Boulevard Descartes, 77 454 Marne-la-Vallée Cedex 2, France*

We use a rigid axisymetric microfluidic flow focusing device to produce monodisperse bubbles, dispersed in a surfactant solution. The gas volume fraction of the dispersion collected out of this device can be as large as 90%, demonstrating that foam with solid-like viscoelastic properties can be produced in this way. The polydispersity of the bubbles is so low that we observe crystallization of our foam. We measure the diameter of the bubbles and compare these data to recent theoretical predictions. The good control over bubble size and foam gas volume fraction shows that our device is a flexible and promising tool to produce calibrated foam at a high flow rate.

**Introduction**

Foams are dispersions of bubbles in an aqueous surfactant solution where neighbouring bubbles touch each other and form a jammed solid-like close packing (1). Only if the applied stress exceeds a yield stress, a foam will flow (2). These properties, combined with the small weight and large specific surface, give rise to a large variety of applications (3), ranging from mineral production to the synthesis of nanoparticles (4) and medical drug delivery (5). There are various techniques for producing aqueous foams, but most involve mixing of the two phases in bulk processes (1);(3). With these methods, the rate of production is high so that industrial quantities can easily be obtained. However, little control over the formation of individual bubbles is available and their size distribution is broad. Monodisperse foams are of interest since they should be particularly stable. Indeed, an aging mechanism of foam, known as coarsening, is driven by the difference of Laplace pressure between neighbouring bubbles. This difference is considerably reduced if the bubbles are monodisperse rather than polydisperse (6). Microfluidics, which allows the control of liquid flow on microscales, offers a new means to produce complex fluids such as monodisperse foams and emulsions (7); (8); (9); (10). In particular, the flow focusing method where two fluids are forced to flow simultaneously through a small orifice, is extremely efficient (11); (12); (13); (14); (15); (16). Depending on flow rate, there is a jetting or a dripping regime, analogous to what is observed in a dripping faucet (18);(19). In the dripping regime, bubbles with a polydispersity index smaller than 2% can be obtained (13). Still, the usefulness of this technique so far appeared to be limited since in most of the cited studies the maximum ratio of gas to liquid flow rates for which the production of monodisperse drops or bubbles has been reported is only of the order of 0.6. This value is equivalent to a gas volume fraction of 40%, far below the fraction of 64% where randomly packed bubbles start touching each other. Only recently, a ratio of gas to liquid flow rate larger than one has been reported (20), but in this case the rate of foam production was very small. This situation motivates the present study of a flow focusing device that can yield foam with gas volume fractions up to 90% at a high rate. To gain physical insight about the bubble formation, we explore the role of the coflowing geometry and show that over a large range of gas to liquid flow rate ratios, the bubble diameter follows a scaling law if it is normalized by the cubic root of the volume of the device orifice, as predicted by a recent model (13).

**Microfluidic device and chemicals**

Our microfluidic device is inspired by the three dimensional flow focusing experiment described in (15). As sketched in Fig 1A, it is made of a cylindrical glass capillary tube fitted very closely into a square glass capillary tube so that good alignment is ensured. The end of the inner tube has a constricted shape obtained by melting it with a hot flame. With this method, we obtain axisymetric constrictions with minimum diameters $b$ ranging from 50 to 350 μm and different lengths that we measure precisely. As we will show, exploring a wide range of aspect ratios is an advantage, since the influence of the constriction volume $\Omega$ and diameter $b$ can be evidenced independently. We measure $\Omega$ using pictures of the constrictions of each device (Fig. 1B). For each picture, we fit the diameter of the constriction as a function of the distance $z$ along the axis of symmetry by a polynomial. $\Omega$ is deduced by integrating over the part of the constriction where its diameter is smaller than 1.5 $b$. The calculated volume $\Omega$ depends only weakly on the choice of the coefficient multiplying $b$ in this criterion since at both ends of the constriction, the diameter varies rapidly as a function of $z$. As shown in Fig. 1A, bubbles are formed by injecting the solution and the gas from the two sides of the square capillary tube and forced to flow through the constriction at the entrance of the inner cylindrical tube. Here, a dispersion of bubbles is formed that leaves the device through the cylindrical tube so that it can be collected. The length $L$ of the collection tube is 5 cm and its radius $R$ is 400 μm. Using a digitally controlled syringe pump (Kd Scientific) the liquid phase is injected into the corners of the square tube at a flow rate denoted as $Q_l$, ranging from 1 to 400 mL/hr. The gas is supplied from a pressurized tank at a pressure $p_g$. For each experiment, we impose $p_g$ and measure the gas volume flow rate at this pressure using a bead-flowmeter. We consider bubbles collected at the exit of the device under the atmospheric pressure $p_o$ and in thermal equilibrium with the surrounding liquid. Thermal equilibrium is important because the gas undergoes an endothermic expansion as the pressure drops from $p_g$ to $p_o$. Furthermore, because the gas is compressible, the total volume of the collected bubbles increases at a rate denoted $Q_g$ larger than the one indicated by the flow meter by a factor $p_g / p_o$, taking into account the isothermal expansion. The obtained values of $Q_g$ are in the range 40 to 500 mL/hr. Note that the Laplace pressure due to the tension of the gas liquid interfaces is negligible for the range of bubble sizes produced in our device.

To observe the bubbles produced in the flow-focusing device right after the constriction, we use a high shutter speed video camera (Marlin) placed above the transparent device. We also collect bubbles at the exit of the device and measure their diameter at the atmospheric pressure $D(p_o)$ using a Nikon microscope. The two immiscible phases used in our experiment are nitrogen gas and an aqueous solution containing a surfactant TetradecylTrimethylAmmonium Bromide (0.936 % g/g), dodecanol (6.24 $10^{-4}$ % g/g) and glycerol (50 % g/g). All the chemicals were purchased from Aldrich and used as received. The concentrations are expressed as the weight of chemicals divided by the total weight of the solution. Water was purified by using a Millipore Milli-Q filtration system. Using the Wilhelmy plate method, the surface tension of the foaming solution was found to be 37.5 ± 0.5 mN/m at a temperature of 21°C and its viscosity $\eta$ is 5 mPa.s.

**Experimental results**

As in previous flow focusing experiments (12); (13); (14); (15); (16), we observe dripping, which produces highly monodisperse bubbles within the constriction of the collection tube as illustrated in Fig. 2. In our axisymmetric geometry, the break-up of the gas thread occurs in three stages. First, the gaseous thread enters the constriction (Fig. 2 A,B), then it proceeds through it until it reaches the collection tube where a bubble is inflated (Fig. 2 C-E). Finally, because the large bubble blocks the exit of the orifice, the liquid radially squeezes the gaseous thread and pinches it off (Fig. 2 F). The large set of monodisperse bubbles made by our device and shown in Fig. 3A demonstrates that the production process is highly reproducible. We observe that up to a depth of three layers inside the sample the bubbles are crystallized into a hexagonal close-packing. However, due to strong scattering of light by the gas liquid interfaces, we cannot tell whether the entire foam has such a structure.

To check whether any bubbles break at the exit of the device and release their gas into the atmosphere, we compare $Q_g$ and $Q_l$ to the gas volume fraction $\Phi_g$ of the dispersion collected at the output of the device. For different $p_g$ and $Q_l$, we fill a container of calibrated volume and determine the gas volume fraction $\Phi_g$ by weighing it. These data, plotted in Fig. 3B as a function of $Q_g / Q_l$, follow the prediction derived from volume conservation:

$$\Phi_g = \frac{Q_g}{Q_g + Q_l} \tag{1}$$

The good agreement demonstrates that the bubbles are not destroyed at the exit of the microfluidic device.

Furthermore, our rigid device allows gas flow rates as high as 400 mL/hr to be injected, 2 orders of magnitude larger than in previously described PDMS based devices (13). Combined with $Q_l$ = 50 mL/hr, we are thus able to produce and collect foam with $\Phi_g$ = 90%. Previous authors have characterized the bubble dispersions inside their devices by i) a gas volume fraction $\Phi_{ch}$ defined as the ratio of the volumes occupied by the moving bubbles to the volume of the outlet channel and ii) a gas volume fraction measured right after the constricted orifice $\Phi_{or}$ and defined as the ratio of the gas flow rate to the total flow rate at the pressure $p_g$. While $\Phi_{ch}$ is reported between 0.5 and 0.9, $\Phi_{or}$ is only comprised between 0.1 and 0.4. $\Phi_{or}$ is always smaller than $\Phi_{ch}$ due to accumulation and rapid flow of liquid in the corners of square outlet channels (13); (20).

In this paper, we focus on the gas volume fraction $\Phi_g$ of the collected dispersion measured at the atmospheric pressure which is relevant in many applications. We obtain $\Phi_g$ between 0.3 and 0.9 for a production rate $Q_l + Q_g$ ranging up to 500 mL/h. Previous authors either report values of $\Phi_{or}$ corresponding to much lower gas volume fractions $\Phi_g < 0.6$ (13), or a very small rate of dry foam production (20). Besides, with $\Phi_g = 0.9$, the regime of jammed close packed bubble dispersions that are called foams is clearly reached. We recall that the close packing fraction of monodispersed spheres is 0.74 for a face centered cubic packing and 0.64 for a random packing. Therefore, the high flow rates in our device allow up to 450 mL/h of monodisperse dry foam to be produced, as required in many applications and fundamental studies.

**Hydrodynamic resistance of the outlet flow**

We now discuss the flow in the collection tube which has a major impact on the hydrodynamic resistance of the device as well as on the bubble formation process itself. This flow, downstream of the constriction where the bubbles are produced, distinguishes our foam generator from previously described microfluidic devices where bubbles are either directly released into a liquid filled tank (14) or confined in channels of such a small height that every bubble touches the wall (13);(20);(21). In the latter case, the pressure drop along the collection tube is governed by the slip of Plateau borders on the channel wall, leading to a non linear relation between the applied pressure and the gas flow rate $p_g \propto Q_g^\beta$ with $\beta$ smaller than one (20). Various studies, going back to the pioneering work of Bretherton (22) have discussed the origin of such behavior. However, in our case, we expect different behavior, since the diameter of the cylindrical collection tube is wide enough to contain many bubbles side by side.

To study the foam flow in the collection tube driven by the applied gas pressure $p_g$ experimentally, we measure $Q_g$ as a function of $p_g$ in one of our devices. These data shown in Fig. 4A are obtained for a constant ratio $Q_g/Q_l$, and hence for a constant gas volume fraction $\Phi_g$ (cf. Eq. (1) and Fig. 3). The data presented in Fig. 4A follow a power law with an exponent of $1.18 \pm 0.06$. Fig 4B shows that the relationship between $p_g$ and $Q_g$ depends only weakly on the gas volume fraction $\Phi_g$. We also observe that all the data shown in Fig 4B obtained at different pressures collapse on the same curve. This allows us to derive an equivalent hydrodynamic resistance of the channel that does not depend on $\Phi_g$ and that is of the order of 8 Bar.s/mL.

To understand the flow patterns that can arise in a channel, we recall that foams are yield stress fluids (2). They can be solid-like or liquid-like, depending on the applied stress. If the applied stress is increased beyond a threshold value that is called the yield stress, the foam undergoes a bulk shear flow. If the applied stress is below the yield stress, there is no relative motion between neighbouring bubbles and this regime is called plug flow. Elementary arguments show that for a steady flow, driven by a pressure $p_g$ in a cylindrical tube of length $L$, the shear stress $\tau(r)$ varies with the distance from the axis of symmetry r as $\tau(r) = p_g \, r/2L$. Using the typical dimensions of our experimental devices, we find that $\tau$ lies between 10 and 240 Pa. Moreover, in view of an empirical expression given in the literature depending on bubble size, surface tension, and gas volume fraction (2), the yield stress $\tau_y$ is comprised between 20 and 300 Pa for all of the foams produced in our experiments. Since the ranges of variation of $\tau_y$ and $\tau$ overlap, we expect in the collection tube a mixed regime with a core undergoing plug flow surrounded by a zone of shear bulk flow. The observed relation between the applied pressure and the gas flow rate $p_g \propto Q_g^{1.18}$ is therefore due to a complex non-Newtonian flow which analysis is beyond the scope of this paper.

**Diameter of the collected bubbles**

Now, we consider the diameter of the produced bubbles. With our four microfluidic devices, monodisperse bubbles with diameter ranging from 50 to 900 µm are produced. Fig. 5A shows the diameter of the bubbles at the atmospheric pressure $D(p_o)$, as a function of $Q_g/Q_l$. For each given constriction diameter, we fit the data with a power law $D(p_o) = D°(Q_g/Q_l)^\alpha$. In Table 1 the parameters $\alpha$ and $D°$ are reported for each of the devices. We observe that the coefficient $D°$ increases with the volume $\Omega$ of the device. To explain the dependency of the bubble diameter with $Q_g/Q_l$, we briefly recall two different theoretical models proposed in the literature (13); (17) The first deals with flows where the Reynolds number in the liquid is high (17). Based on a discussion of the unsteady and convective terms of the Navier-Stokes equation the bubble diameter is found to scale as $b \, (Q_g/Q_l)^{0.4}$, in good agreement with the author's experimental study (16);(17). An alternate model predicts a linear scaling of the bubble volume with $Q_g/Q_l$ and explains the bubble formation process (13) as well as its excellent repeatability as follows: first the gaseous thread advances into the constriction until it reaches the collection tube where a bubble is inflated. As soon as this bubble is large enough to block the exit of the constriction, the liquid will squeeze the gaseous thread. It finally reaches a size for which it is unstable and breaks rapidly. The progression of this collapse and consequently the rate of bubble formation are controlled by the flow rate of supplied liquid that squeezes the gaseous thread. Because of the confinement, the liquid/gas interface is stable during most of the break-up; Therefore, the collapsing interface proceeds through a series of quasistationary states. The last step of the collapse is not as reproducible as the previous ones, but since it is fast compared to the total break-up time, the process as a whole is periodic. Moreover, this model predicts that the bubble formation period is given by the time required to fill up the constriction by the liquid flow. Assuming conservation of the gas volume, this subsequently determines the diameter $D(p_o)$ of a bubble, at the outlet of the device:

$$D(p_o) = 2 \left(\frac{3}{4\pi} \Omega\right)^{1/3} \left(\frac{Q_g}{Q_l}\right)^{1/3} \qquad (2)$$

Therefore, the bubble diameters should collapse onto a master curve if they are scaled by $\Omega^{1/3}$, in contrast to the first model, which predicts a scaling with $b$ (13); (17). We find that scaling the bubble diameter by $\Omega^{1/3}$ rather than by $b$ yields the best masterplot; the sum of the least square differences between the estimated points and the observed values is equal to 8.6 for data normalized by $\Omega^{1/3}$ whereas it is equal to 88.1 for the data normalized by $b$. Fig. 5B shows the collapse of the whole set of our data scaled by $\Omega^{1/3}$. Moreover, we find that the measured prefactors $D°/\Omega^{1/3}$ are comprised between 1.0 and 1.4 (cf. Table 1), consistent with the value 1.24 predicted by equation 2. Finally, the average of the power law exponents obtained for our devices is 0.38, close to the value 0.33 expected from equation 2. Therefore, the second model seems to provide the best physical description of our device: it not only accounts for the monodispersity of the bubbles but also predicts the scaling of the bubble diameter with the volume of the constriction. Still, the discrepancy between the observed and the predicted exponents of the power law shows that the physics of bubble production is not yet fully captured.

**Conclusion**

Our experiments demonstrate how a rigid axisymetric flow focusing device can produce large quantities of monodisperse foam with bubble diameters ranging from 50 to 900 µm and a gas volume fraction up to 90%. We observe over a broad range of foam gas volume fractions and bubble sizes that the hydrodynamic resistance for the gas flow driving the foam in the collection tube depends only weakly on the applied pressure. In this context, further investigations of the interplay between wall slip and bulk flow as well as of the impact of the constriction on global flow resistance of the device would be of great interest. Finally, we find that if the compressibility of the gas is taken into account, the gas to liquid flow rate ratio governs both the foam gas volume fraction and the bubble diameter. The latter relation is found to obey a scaling law that extends over a range of flow rate ratios two orders of magnitude larger than previously reported. Furthermore, our experiments strongly suggest that the bubble diameter formed within the microfluidic device scales with the cubic root of the constriction volume rather than with the constriction width, confirming a previous conjecture. All these different findings will help to make progress towards a quantitative understanding of the physics of flow focusing foam generators, and to make these devices even more useful in applications.

ACKNOWLEDGEMENT

E.L. thanks Prof D. A. Weitz and Prof. H. A. Stone for having brought this subject to her mind during her post-doctoral fellowship. E.L is also indebted to A. S. Utada for fructuous discussions. We thank D. Hautemayou and J. Laurent for their technical help.

# REFERENCES

(1) Weaire, D. and Hutzler S. (1999). The Physics of Foams. Oxford, Oxford University Press.
(2) Höhler, R. And Cohen-Addad S. (2005). "Rheology of liquid foams." Journal of Physics: Condensed Matter **17**: R1041-R1069.
(3) Khan, S. A. and Prud'homme R. (1996). Foams. New York, Marcel Dekker.
(4) Bala, T., Sankar, C.R., Baidakova, M., Osipov, V., Enoki, T., Joy, P.A., Prasad, B.L.V., Sastry, M. (2005). "Cobalt and magnesium ferrite nanoparticles: preparation using liquid foams as templates and their magnetic characteristics." Langmuir **21**(23): 10638-43.
(5) Sadick, N. S. (2005). "Advances in the treatment of varicose veins: ambulatory phlebectomy, foam sclerotherapy, endovascular laser, and radiofrequency closure." Dermatologic Clinics **23**(3): 443-455.
(6) Gañán-Calvo, A.M., Fernández, J.M., Oliver, A.M., Marquez, M. (2004). "Coarsening of monodisperse wet microfoams." Applied Phys. Lett. **84**(24): 4989-4990.
(7) Tice, J.D., Song, H., Lyon, A.D., Ismagilov, R.F. (2003). "Formation of Droplets and Mixing in Multiphase Microfluidics at Low Values of the Reynolds and the Capillary Numbers." Langmuir **19**: 9127-9133.
(8) Okushima, S., Nisisako, T., Torii, T., Higuchi, T. (2004). "Controlled production of monodisperse double emulsions by two-step droplet breakup in microfluidic devices." Langmuir **20**: 9905-9908.
(9) Atencia, J. and Beebe D. (2005). "Controlled microfluidic interfaces." Nature **437**: 650 - 655.
(10) Nisisako, T., Okushima, S., Torii, T. (2005). "Controlled formulation of monodisperse double emulsions in a multiple-phase microfluidic system." Soft Matter **1**: 23-27.
(11) Gordillo, J.M., Gañán-Calvo, A.M., Pŕez-Saborid, M. (2001). "Monodisperse microbubbling: absolute instabilities in coflowing gas-liquid jets." Physics of fluids **13**(12): 3839-3842.
(12) Anna, S.L., Bontoux, N., Stone, H.A. (2003). "Formation of dispersions using ``flow focusing'' in microchannels." Applied Phys. Lett. **82**(3): 364-366.
(13) Garstecki, P., Gitlin, I., Diluzio, W., Whitesides, G.M., Kumacheva, E., Stone, H.A. (2004). "Formation of monodisperse bubbles in a microfluidic flow-focusing device." Applied Phys. Lett. **85**(13): 2649-2651.
(14) Gordillo, J.M., Cheng, Z., Ganan-Calvo, A.M., Márquez, M., Weitz, D.A. (2004). "A new device for the generation of microbubbles." Physics of Fluids **16**(8): 2828-2834.
(15) Utada, A.S., Lorenceau, E., Link, D.R., Kaplan, P.D., Stone, H.A., Weitz, D.A. (2005). "Monodisperse Double Emulsions Generated from a Microcapillary Device." Science **308**: 537-541.
(16) Gañán-Calvo, A.M., Gordillo, J.M. (2001). "Perfectly monodisperse microbubbling by capillary flow focusing ." Phys. Rev. Lett. **87**: 2745011-2745014
(17) Gañán-Calvo, A.M. (2004). " Perfectly monodisperse microbubbling by capillary flow focusing: An alternate physical description and universal scaling." Phys. Rev. E **69**: 027301-1-027301-3
(18) Clanet, C., Lasheras, J.C. (1999). " Transition from dripping to jetting." J. Fluid Mech. **383**: 307-326.
(19) Ambravaneswaran, B., Subramani, H.J., Phillips, S.D., Basaran, O.A. (2004). " Dripping-jetting transitions in a dripping faucet." Phys. Rev. Lett. **93**(3): 034501-1.
(20) Raven, J.-P., Marmottant, P., Graner, F. (2006). "Dry microfoams: Formation and flow in a confined channel" European Physical Journal B **51**: 137-143

(21) Cubaud, T., Ho, C.-M. (2004). "Transport of bubbles in square microchannels." <u>Physics of Fluids</u> **16**(12): 4575-4585.

(22) Bretherton, F.P. (1961). "The motion of long bubbles in tubes." <u>J. Fluid Mech.</u> **10**: 166 - 188.

TABLE

| Device | α | $D°$ (μm) | $D°/\Omega^{1/3}$ | $\Omega$ (nL) | $b$ (μm) | $\Omega^{1/3}$ (μm) |
|---|---|---|---|---|---|---|
| 1 | 0.36±0.01 | 403±4 | 1.01 | 63.0 | 345±5 | 398±20 |
| 2 | 0.38±0.02 | 322±6 | 1.35 | 13.3 | 148±5 | 237±16 |
| 3 | 0.39±0.02 | 154±5 | 0.85 | 6.0 | 192±5 | 182±15 |
| 4 | 0.38±0.2 | 57±20 | 1.1 | 0.1 | 50±5 | 52±10 |

Table 1: Characteristics of the bubbles and geometrical properties of the four different microfluidic devices. See the text for the definition of the notation.

FIGURE CAPTIONS

Figure 1 : A) Schematic of the coaxial microfluidic device. The collection tube (shown on the right) is a cylindrical capillary of 400 μm radius and 5 cm length. It is constricted on the side that is inserted into the square tube. The inner dimension of the square tube is 1 mm; This matches the outer diameter of the untapered region of the collection tube. The diameter of the narrow orifice in the collection tube lies in the range 50 to 350 μm. B) Photograph of the constriction of device 4. Its profile is fitted by a polynomial function, represented by the black lines.

Figure 2 : Formation of a bubble in the constriction. Photographs A to F correspond to successive snapshots. Gas and liquid appear respectively as dark and bright regions. $b$=148 μm, $Q_g$ =75 mL/hr, $Q_l$ = 50 mL/hr.

Figure 3 : A) Top view of hexagonally close packed monodisperse bubbles (diameter: 250 μm) collected out of a device with constriction diameter 148 μm. B) Foam gas volume fraction as a function of the gas to liquid flow rate ratio. The continuous line is the gas volume fraction predicted from volume conservation (equation 1). The dashed line indicates the gas volume fraction of the close packing of monodisperse spherical bubbles.

Figure 4 : A) Gas pressure versus gas flow rate for a constant gas volume fraction of 70 %, obtained using device 2. The continuous line is a fitted power law with an exponent 1.18. B) Global hydrodynamic resistance of the microfluidic device 2 versus liquid to gas flow rate ratio ( $b$=148 μm). The equation of the straight line is -0.4 $\phi_g$ +8.3.

Figure 5 : A) Bubble diameter measured at the atmospheric pressure versus the gas to liquid flow rate ratio for four different devices whose characteristics are given in Table 1. Each data set is fitted by a power law corresponding to a straight line (parameters: see Table 1). B)

Bubble diameter measured at the atmospheric pressure scaled by the constriction volume $\Omega$ to the power 1/3, versus the gas to liquid flow rate ratio. The symbols correspond to the 4 different devices as in A). The straight line corresponds to the average power law 1.1 $(Q_g/Q_l)^{0.38}$
.

FIGURES

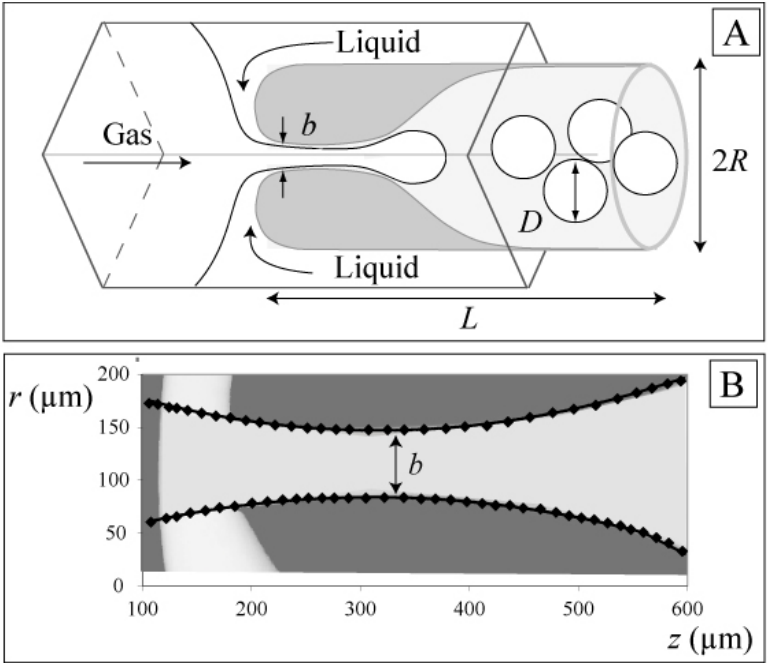

Fig. 1, Lorenceau et al.

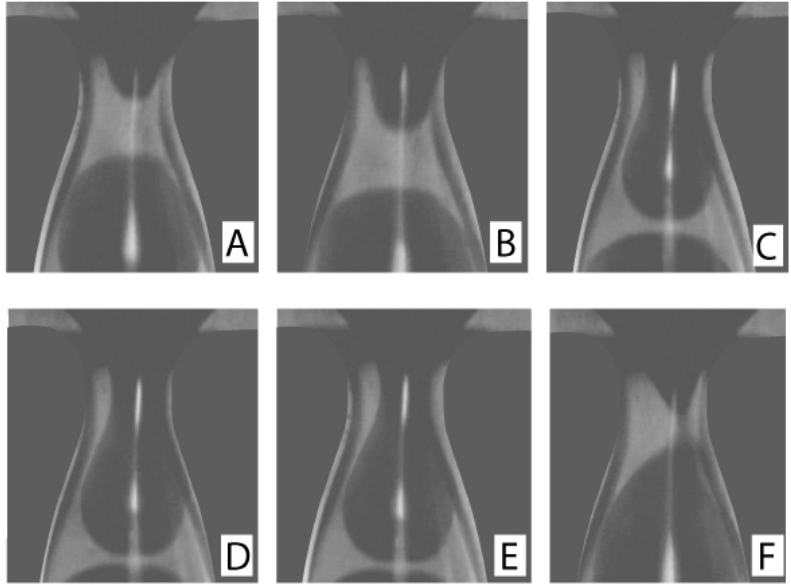

Fig. 2, Lorenceau et al.

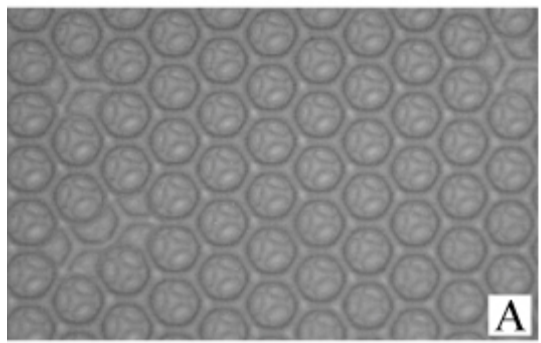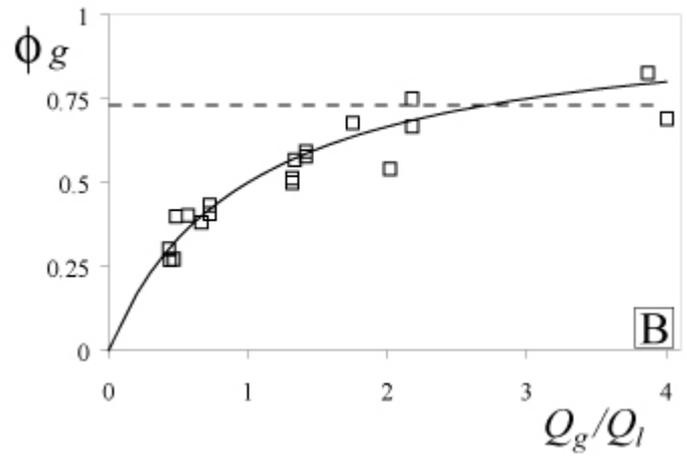

Fig. 3, Lorenceau et al.

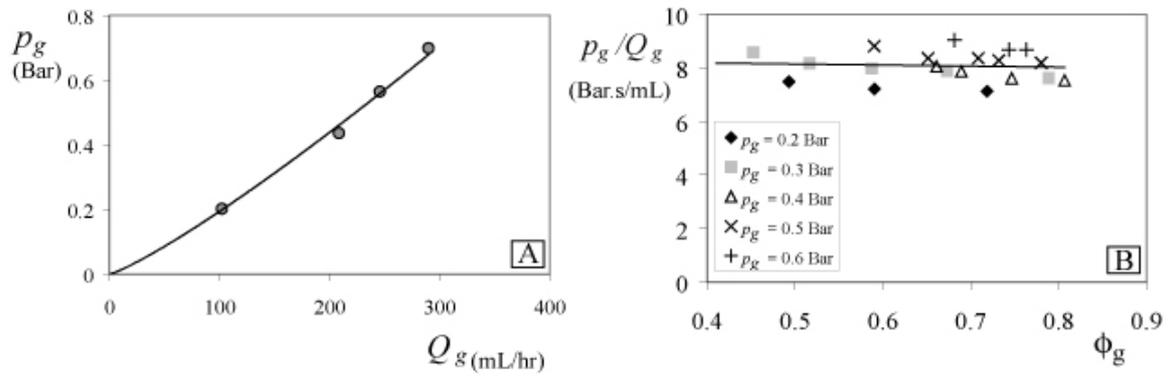

Fig. 4, Lorenceau et al.

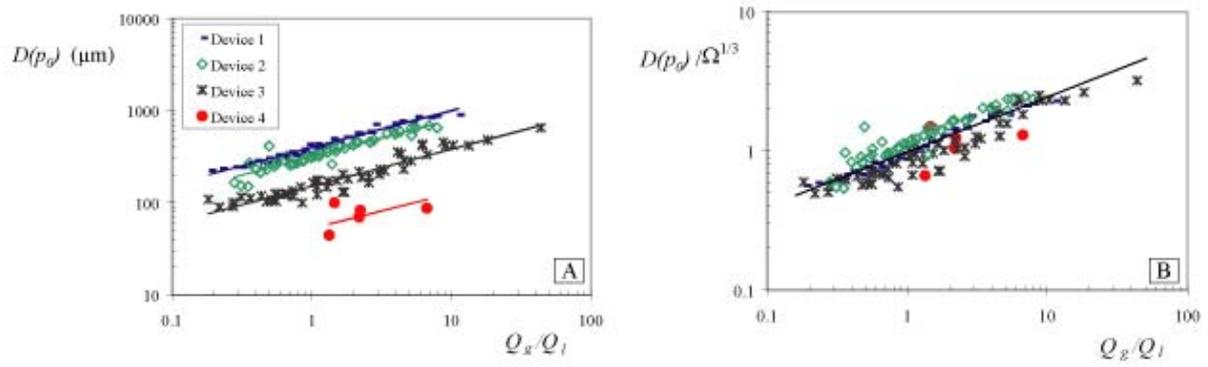

Fig. 5, Lorenceau et al.